\documentclass[twocolumn,showpacs,preprintnumbers,amsmath,amssymb]{revtex4}
\usepackage{amssymb}
\usepackage{amsfonts}
\usepackage{mathrsfs}
\usepackage{amsmath}
\usepackage{graphicx}
\usepackage{dcolumn}
\usepackage{bm}
\usepackage{hyperref}

\begin{document} 
\title{Robust quantum repeater with atomic ensembles against phase and polarization instability}
\author{Ming Gao$^{1,2}$}
 \author{Lin-Mei Liang$^2$}
 \author{Cheng-Zu Li$^2$}
\author{Xiang-Bin Wang$^1$}
 \email{xbwang@mail.tsinghua.edu.cn}
\affiliation{1.Department of Physics, Tsinghua University, 100084,
China.\\ 2.Department of Physics, National University of Defense
Technology, Changsha, 410073, China.}
\date{\today}
\begin{abstract}

We propose an alternative scheme for quantum repeater without phase
stabilization and polarization calibration of photons transmitted
over long-distance channel. We introduce time-bin photonic states
and use a new two-photon interference configuration to robustly
generate entanglement between distant atomic-ensemble-based memory
qubits. Our scheme can be performed with current experimental setups
through making some simple adjustments.
 \end{abstract}

\pacs{03.67.-a, 03.67.Mn, 03.67.Pp} 

\maketitle

\section{\label{sec:Introd}Introduction}
The concept of quantum repeater \cite{BDCZ}was introduced for
long-distance quantum communication in order to overcome the
problems caused by inevitable photon loss in the transmission
channel. Generating distant entanglement is a crucial ingredient of
a quantum repeater protocol. In 2001, Duan, Lukin, Cirac and Zoller
(DLCZ) proposed a original scheme \cite{DLCZ} to use
atomic-ensembles and linear optics in which robustly generating
entanglement over long distances can be achieved. Motivated by the
DLCZ protocol, much experimental effort\cite{Matsukevich,Chou,Chen}
has been made in the last few years.

However, it has been shown that DLCZ protocol requires severe phase
stability since entanglement generation and entanglement swapping in
the protocol depend on single-photon Mach-Zehnder type
interferences\cite{BZhao}, and this problem make its experimental
realization extremely difficult by the current technology. Hence a
novel scheme\cite{BZhao} based on DLCZ scheme was proposed to use
phase-insensitive two-photon quantum interference which dramatically
relax this phase stability requirements. A latest
experiment\cite{zhshyuan} has primarily demonstrated the scheme over
a distance of 300 meters. Meanwhile, a fast and robust approach
\cite{otherwork} was proposed and also can solve the phase stability
problem as long as the entanglement generation is performed locally,
and they gave a detailed comparisons between the schemes. In
addition, there are another novel schemes
\cite{efficiency1,efficiency2} proposed to improve the efficiency of
quantum repeaters.

These protocols greatly stimulate experimental implementations of
quantum repeater, but there are still some problems need to be
solved, such as reliable transmission of photon's polarization
states over noisy channel. Since the photon interferences rely on
the polarization states, the ability to maintain photonic
polarizations is indispensable in the process of distant
entanglement generation or swapping. Most of the time, optical
fibers are used as photon transmission channel. Due to the fiber
birefringence, the photonic polarizations will be changed randomly
\cite{Gisin}. Experimentally, active feedback compensation could be
applied to solve this problem\cite{CZPeng}, but it is efficient only
when the thermal and mechanical fluctuations are rather slow.
Furthermore, even though polarization compensations can be used
efficiently, imperfect shared reference frame (SRF) for polarization
orientation may cause some errors. It is difficult to correct this
kind of errors since establishing a perfect SRF requires infinite
communication\cite{Rudolph}. Due to these reasons, it would be
better to have a quantum repeater scheme with inherent polarization
insensitivity.

In this paper, we propose an alternative approach to create distant
entanglement between atomic-ensembles. In our scheme neither the
phase stabilizing nor the polarization calibrating is needed for
photons transmitted over long distance. Through introducing time-bin
photonic states and using a new two-photon interference
configuration, we make only the unchanged part of initial
polarization states contribute to the desired results. Combined with
local entanglement swapping for entanglement connection, our scheme
can be used to implement a robust quantum repeater.

\section{\label{sec:ENG}Entanglement generation}
\begin{figure}
\scalebox{0.95}{\includegraphics[width=\columnwidth]{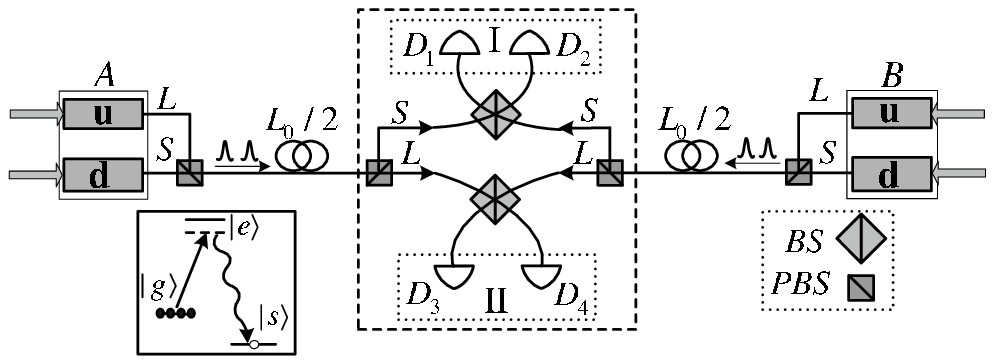}}
\caption{\label{fig:1}Schematic of our scheme. At each node, the
Stokes photons generated from the two atomic ensembles have
orthogonal polarizations, i.e.$|H\rangle$ and $|V\rangle$. The $PBS$
transmits $|H\rangle$ and reflects $|V\rangle$. Before leaving the
node, the photon wave packets with different polarizations
experience different paths and then travel to the middle point where
they experience a same path difference but with opposite
arrangement. There are two detection zones: one with the single
photon detectors $D_{1,2}$ and the other with $D_{3,4}$. A
coincidence count between the detectors at either detection zone,
e.g. $D_1$ and $D_3$, will project the four atomic ensembles into a
complex entangled state up to a local unitary transformation. The
inset shows the ensemble atom energy level.}
\end{figure}

Here optical thick atomic ensemble, which includes $N$ atoms with
$\Lambda$ level structure (see inset of Fig.\ref{fig:1}), is used as
quantum memory. Each atomic ensemble is illuminated by a short,
off-resonant write pulse that induces a spontaneous Raman process.
This process will produce a forward-scattered Stokes light and a
collective atomic excitation state\cite{DLCZ}. The photon-atom
system can be described as(neglecting the higher-order terms)
\begin{eqnarray} \label{eq:initial}
[1+\sqrt{\chi}S^{\dag}a^{\dag}+\frac{\chi}{2}(S^{\dag}a^{\dag})^2]|vac\rangle,
\end{eqnarray}
where $|vac\rangle$ denotes that all the ensemble atoms are in the
ground state $|g\rangle$ and the Stokes light in the vacuum state,
$a^{\dag}$ is the creation operator of the Stokes light, and the
collective atomic excitation is defined by
$S^{\dag}=\frac{1}{\sqrt{N}}\sum_{i=1}^N|s\rangle_{i}\langle g|$.
The excitation probability $\chi\ll1$ can be achieved by
manipulating the write laser pulse.

At each communication node, two atomic ensembles are used to encode
a memory qubit(see the Fig.\ref{fig:1}). The two ensembles are
excited simultaneously by write laser pulses, and the Stokes photons
generated from them have orthogonal polarization states, i.e.
$|H\rangle$ and $|V\rangle$, which denote horizontal and vertical
linear polarization, respectively. The state $|H\rangle$ propagates
through a short path ($S$) and the state $|V\rangle$ goes through a
long path ($L$), and are combined at a polarization beam splitter
($PBS$) which transmits $|H\rangle$ and reflects $|V\rangle$.
Therefore photon wave packets with different polarizations
correspond to different time bins. As long as the path difference is
less than the photons coherence length (about $3m$ or more for
photons generated from atomic ensembles\cite{Eisaman}), after the
PBS the memory qubit is effectively entangled with the polarization
and the time-bin states of the emitted Stokes photons. The state of
the atom-photon system can be written as
\begin{eqnarray}
\begin{split}
|\psi\rangle&=\{1+\sqrt{\chi}(S_u^{\dag}a_{V,L}^{\dag}+S_d^{\dag}a_{H,S}^{\dag})+\frac{\chi}{2}[(S_u^{\dag}a_{V,L}^{\dag})^2\\
&+2S_u^{\dag}a_{V,L}^{\dag}S_d^{\dag}a_{H,S}^{\dag}+(S_d^{\dag}a_{H,S}^{\dag})^2]\}|vac\rangle,
\end{split}
\end{eqnarray}
where the subscripts $u$ and $d$ are used to distinguish the two
ensembles, and $a^{\dag}_{V,L}$($a^{\dag}_{H,S}$) denotes the
creation operator of the Stokes photon with vertical (horizontal)
polarization passing through the long (short) path.

Assume two neighboring communication nodes, denoted by $A$ and $B$,
are connected through certain transmission channel such as optical
fiber with a distance of $L_0$. At the middle point between the two
nodes, the Stokes photons are firstly directed to PBS. Due to
channel noise or imperfect SRF, the polarization states of photons
arrived at the middle point have been changed, and as long as the
time intervals between the subsequent transmitted photon wave
packets are small, it is reasonable to assume that the polarization
changes are the same to them. After the PBS, there are also a long
path and a short one whose difference is the same as the one at each
node. The polarization transformation during this process are
usually considered as a unitary transformation\cite{Gisin}, so the
evolution of the photonic polarization components can be described
as:
\begin{eqnarray}\label{eq:photon evolution}
\begin{split}
a^{\dag}_{H,S}&\underrightarrow{{\quad}noise{\quad}}\cos{\theta}a^{\dag}_{H,S}+e^{i\varphi}\sin{\theta}a^{\dag}_{V,S}\\
&\underrightarrow{{\quad}PBS{\quad}}\cos{\theta}a^{\dag}_{H,SL}+e^{i\varphi}\sin{\theta}a^{\dag}_{V,SS},\\
\end{split}
\end{eqnarray}
\begin{eqnarray}\label{eq:photon evolution}
\begin{split}
a^{\dag}_{V,L}&\underrightarrow{{\quad}noise{\quad}}-e^{-i\varphi}\sin{\theta}a^{\dag}_{H,L}+\cos{\theta}a^{\dag}_{V,L}\\
&\underrightarrow{{\quad}PBS{\quad}}-e^{-i\varphi}\sin{\theta}a^{\dag}_{H,LL}+\cos{\theta}a^{\dag}_{V,LS},
\end{split}
\end{eqnarray}
where $\theta$ and $\varphi$ are random noise parameters. Obviously,
there are four time bins $SS,SL,LS$ and $LL$ for each Stokes photon,
and $SL,LS$ are the same time bin. Now the atom-photon system can be
described as:
\begin{eqnarray} \label{eq:state at M}
 |\psi\rangle=|vac\rangle+|\psi_{LS,SL}\rangle+|\psi_{SS,LL}\rangle+|\psi_{cross}\rangle,
\end{eqnarray}
with
\begin{eqnarray} \label{eq:time bins}
\begin{split}
|\psi_{LS,SL}\rangle&=\{\cos{\theta}\sqrt{\chi}(S_u^{\dag}a_{V,LS}^{\dag}+S_d^{\dag}a_{H,SL}^{\dag})\\
&+\frac{\chi}{2}\cos^2{\theta}[(S_u^{\dag}a_{V,LS}^{\dag})^2+(S_d^{\dag}a_{H,SL}^{\dag})^2\\
&+2S_u^{\dag}S_d^{\dag}a_{V,LS}^{\dag}a_{H,SL}^{\dag}]\}|vac\rangle,
\end{split}
\end{eqnarray}
\begin{eqnarray} \label{eq:time bins}
\begin{split}
|\psi_{SS,LL}\rangle&=\{\sqrt{\chi}\sin{\theta}(S_d^{\dag}a_{SS}^{\dag}-S_u^{\dag}a_{LL}^{\dag})\\
&+\frac{\chi}{2}\sin^2{\theta}[(S_u^{\dag}a_{LL}^{\dag})^2+(S_d^{\dag}a_{SS}^{\dag})^2\\
&-2S_u^{\dag}S_d^{\dag}a_{LL}^{\dag}a_{SS}^{\dag}]\}|vac\rangle,
\end{split}
\end{eqnarray}\begin{eqnarray} \label{eq:time bins}
\begin{split}
|\psi_{cross}\rangle&=\chi\sin{\theta}\cos{\theta}[(S_d^{\dag})^2a_{SL}^{\dag}a_{SS}^{\dag}
-(S_u^{\dag})^2a_{LL}^{\dag}a_{LS}^{\dag}\\
&+S_u^{\dag}S_d^{\dag}(a_{LS}^{\dag}a_{SS}^{\dag}-a_{LL}^{\dag}a_{SL}^{\dag})]\}|vac\rangle,
\end{split}
\end{eqnarray}
Here for simplicity the $\varphi$ and the polarization subscripts
are not visibly expressed in the states $|\psi_{SS,LL}\rangle$ and
$|\psi_{cross}\rangle$ without any influences on the subsequent
analysis.

The setup for photon interferences at the middle point (see
Fig.\ref{fig:1}) is that the photons after the $PBS$ are directed
into beam splitters ($BS$) followed by single photon detectors which
are turned on only at the time $SL$($LS$). Therefore the terms only
including time bins $SS$ and $LL$ will not have any contributions to
the detection results and can be safely neglected. To generate
entanglement between the nodes $A$ and $B$, laser pulses excite the
ensembles in both nodes simultaneously, and the whole system is
described by the state $|\psi\rangle_A\bigotimes|\psi\rangle_B$,
where $|\psi\rangle_A$ and $|\psi\rangle_B$ are given by equation
(\ref{eq:state at M}) with all the operators and states
distinguished by the subscript $A$ and $B$. A coincidence count
between the detectors at either detection zone, e.g. $D_1$ and
$D_3$, will project the neighboring memory qubits into a complex
state with contributions from second-order excitations. Note that in
second order in $\chi$, the states $|\psi_{cross}\rangle_{A,B}$ can
just trigger one detector at either detection zone at the time
$SL$($LS$) and consequently have no contribution to the coincidence.
Thus a coincidence count between detectors, for instance $D_1$ and
$D_3$, projects the two memory quits into
\begin{eqnarray} \label{eq:final state}
\begin{split}
&|\psi\rangle_{AB}=\frac{1}{2}[e^{i(\phi+\phi^{\prime})}\cos{\theta}\cos{\theta^{\prime}}(S_{u_A}^{\dag}S_{d_B}^{\dag}
+S_{u_A}^{\dag}S_{d_B}^{\dag})\\
&{\quad}e^{2i\phi}\cos^2{\theta}S_{u_A}^{\dag}S_{d_A}^{\dag}
+e^{2i\phi^{\prime}}\cos^2{\theta^{\prime}}S_{u_B}^{\dag}S_{d_B}^{\dag}]|vac\rangle\,
\end{split}
\end{eqnarray}
where $\theta$($\theta^{\prime}$) and $\phi$($\phi^{\prime}$) are
the noise parameter and the polarization-independent phase that the
photons acquires during the transmission from the node $A$($B$) to
the middle point, respectively. The first part of the state is the
desired maximally entangled state. The second part is unwanted
two-excitation state and can be effectively eliminated by
entanglement swapping. The success probability is on the order of
$O(\chi^2\eta^2e^{-L_0/L_{att}})$, where $\eta$ is the detection
efficiency and $L_{att}$ is the channel attenuation length. Until
now, we have generated the state similar to the one (Eq.(S3) in
Supplementary information of Ref.\cite{zhshyuan}) which was created
in the recent experimental demonstration of quantum repeater over a
distance of 300 meters. It is obvious that the polarization noises
only influence the success probability but have no effect on the
fidelity of the desired state, and the phases $\phi$ and
$\phi^{\prime}$ only lead to a trivial global factor
$e^{i(\phi+\phi^{\prime})}$ on the desired state.

\section{\label{sec:singlequbit}local entanglement swapping for entanglement connection}
\begin{figure}
\scalebox{0.95}{\includegraphics[width=\columnwidth]{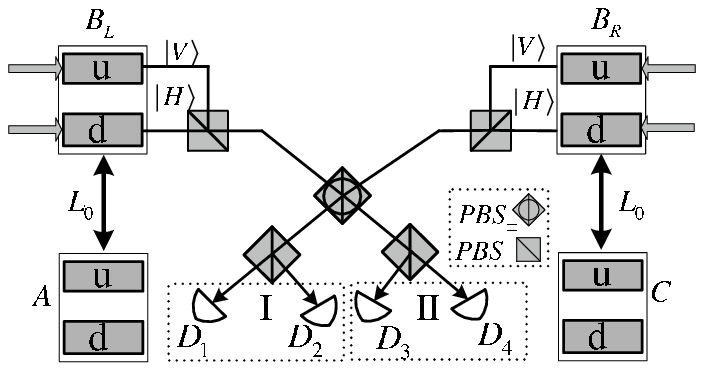}}
\caption{\label{fig:ENW}Set-up for local entanglement swapping.
There are three nodes $A$, $B$ and $C$. At node $B$, the anti-Stokes
photons retrieved from memory qubits $B_L$ and $B_R$ have orthogonal
polarizations $|H\rangle$ and $|V\rangle$. The $PBS$($PBS_{\pm}$)
transmits $|H\rangle$($|+\rangle$) and reflects
$|V\rangle$($|-\rangle$),
where$|\pm\rangle=\frac{1}{\sqrt{2}}|H\rangle\pm|V\rangle$. The
$D_{1,2,3,4}$ are single photon detectors at two detection zones.}
\end{figure}

The entanglement swapping setup is depicted in Fig.\ref{fig:ENW},
which is the same as the one in the schemes\cite{BZhao}. Note that
there is no path difference at this situation. Consider three
communication nodes $A$, $B$ and $C$, and assume that we have
created the complex entangled states (given by Eq.(\ref{eq:final
state})) $|\psi\rangle_{AB_L}$ and $|\psi\rangle_{B_RC}$ both in
($A,B_L$) and in ($B_R,C$), respectively. The memory qubits $B_L$
and $B_R$ at node $B$ are illuminated simultaneously by retrieval
laser pulses. The retrieved anti-Stokes photons are subject to
Bell-state measurement ($BSM$) which is used to eliminate the
two-excitation terms since the arrangement of the $PBS$s is to
identify $|\phi^{\pm}\rangle$at $|+\rangle/|-\rangle$ polarization
basis, where
$|\phi^{\pm}\rangle=\frac{1}{\sqrt{2}}(|+\rangle|+\rangle\pm|-\rangle|-\rangle)$.
Therefore the two-photon states generated from the unwanted
two-excitation terms are directed into the same detection zone and
will not induce any coincidences. In addition, the capability of
distinguishing photon numbers is technically demanding, and the
retrieve efficiency is determined by the optical depth of the atomic
ensembles\cite{Gorshkov}. Taking into account of these
imperfections, the coincidence counts in $BSM$ actually prepare the
memory qubits into a mixed entangled state of the form
$\rho_{AC}=p_2\rho_2+p_1\rho_1+p_0\rho_0$, where the coefficients
$p_2$, $p_1$ and $p_0$ are determined by the retrieval efficiency
and detection efficiency\cite{BZhao}. Here
$\rho_2=|\phi\rangle_{AC}\langle \phi|$ is the maximal entangled
state, where
$|\phi\rangle_{AC}=\frac{1}{\sqrt{2}}(S_{u_A}^{\dag}S_{u_C}^{\dag}+S_{d_A}^{\dag}S_{d_C}^{\dag})|vac\rangle$,
$\rho_0$ is the vacuum state that all the atomic ensembles are in
the ground states, and $\rho_1$ is a maximally mixed state in which
only one of the four atomic ensembles has one excitation.

Although $\rho_{AC}$ still includes unwanted terms, it can be
projected automatically to maximally entangled states in the
entanglement-based quantum cryptography schemes. When implementing
quantum cryptograph via Ekert protocol\cite{Ekert}, only the
coincidence counts between the detectors at two remote nodes are
registered and used for quantum cryptography. Therefore only the
maximally entangled state in $\rho_{AC}$ will contribute to the
experimental results. In this sense, $\rho_{AC}$ is equivalent to
the Bell state $|\phi\rangle_{AC}$.

To implement a quantum repeater protocol, further entanglement
swapping is required. Since the noise parameters only change the
coefficients of the obtained states, the analysis of the
entanglement connection in Ref.\cite{BZhao} can be directly used
here, that is, as long as the excitation probability $\chi$ is small
enough, the contributions from the higher-order excitations can be
safely neglected. The probability to find an desired entangled pair
in the remaining memory qubits is almost a constant and will not
decrease significantly during the entanglement connection process.

\section{\label{sec:conclu} conclusion}
In the existing quantum repeater protocols, the ability to reliably
transfer of photon's polarization is indispensable, but it is not
easy to meet the prerequisite in practice. For this reason, we have
proposed an alternative approach with inherent polarization
insensitivity to generate entanglement between distant communication
nodes. In our scheme, neither the phase stabilizing nor the
polarization calibrating is needed for photons transmitted over long
distance. Through introducing path difference, we make photon wave
packets with different polarizations correspond to different
time-bins, and use the two-photon interference with different
configuration to generate entanglement between remote communication
nodes. Hence only the unchanged part of initial polarization states
can induce the coincidences between detectors, so that the
polarization noise only have influence on the success probability
but the fidelity of the desired states. Consider the function of
atom ensembles as quantum memory for entanglement of a storage time
up to milliseconds\cite{ms}, the capacity of creating high-fidelity
entangled states may be preferred even if at the cost of some
efficiencies. Combined with local entanglement swapping for
entanglement connection, our scheme can be used to implement a
robust quantum repeater. Comparing with the DLCZ scheme and the
scheme of Ref.\cite{BZhao}, the generated entangled states in our
scheme are of higher fidelity.

Finally, it is pointed out that our scheme can be used to perform
remote entanglement swapping after locally generating entanglement,
which would make the quantum repeater scheme more
efficient\cite{efficiency2}. For instance, the local entangled state
(eq.(10) in Ref.\cite{BZhao}) generated via the setup in
Fig.\ref{fig:ENW} may be robustly connected to longer communication
distance by use of our scheme.

\begin{acknowledgments}we thank Cheng-Xi Yang, Lin Yang and Xian-Min Jin  for useful discussions.
X.B.Wang was supported in part by the National Basic Research
Program of China grant No. 2007CB907900, and 2007CB807901, NSFC
grant No. 60725416 and China Hi-Tech program grant No. 2006AA01Z420.
L.-M. Liang was supported by National Funds of Natural Science grant
No. 10504042.
\end{acknowledgments}


\end{document}